\documentclass[aps,pra,preprint,showpacs,showkeys]{revtex4}
\usepackage{amssymb,bm}
\usepackage{graphicx}
\usepackage{amsmath}
\usepackage{epstopdf}
\usepackage{float}
\allowdisplaybreaks

\begin{document}

\title{High-energy electroproduction in an atomic  field}

\author{P. A. Krachkov}\email{peter_phys@mail.ru}
\author{A. I. Milstein}\email{A.I.Milstein@inp.nsk.su}
\affiliation{Budker Institute of Nuclear Physics, 630090 Novosibirsk, Russia}

\date{\today}

\begin{abstract}
The differential cross section of high-energy electroproduction in the electric field of  heavy atoms is derived. The result is obtained with the exact account  of the atomic field by means of  the quasiclassical approximation to the wave functions in the external field. The Coulomb corrections substantially modify  the differential cross section compared with the Born result.  They lead to the azimuth asymmetry in the differential cross section for the polarized incoming electron.  The Coulomb corrections to the total cross section are obtained in the leading logarithmic approximation.
\end{abstract}

\pacs{12.20.Ds, 32.80.-t}

\keywords{electroproduction, photoproduction, bremsstrahlung, Coulomb corrections, screening}

\maketitle

\section{Introduction}
 The process of $e^+e^-$ pair production at   collisions of high-energy electron  with atoms,  which is commonly referred to as  electroproduction or the trident process, is one of the most interesting and important QED processes. This process should be taken into account  when considering  electromagnetic showers in detectors. Electroproduction  is also  important  in some fixed target experiments, see, e.g., dark-photon search experiments \cite{A1,APEX}. In these experiments  electroproduction is the basic irreducible background process.

  The process of electroproduction  has been under consideration for a long time. The earliest  papers are those of Bhabha \cite{Bhabha1,Bhabha2} and Racah \cite{Racah36,Racah37}. In Refs.~\cite{Bhabha1,Bhabha2} calculations were performed with the  use of the Weizs\"acker-Williams approximation (see, e.g., Ref.~\cite{BLP1982}), which allows one  to calculate the total cross section of the  process in the  leading approximation with respect to the  parameter $\ln(\varepsilon/m)$, here  $m$ is the electron mass and  $\varepsilon$ is  the  energy  of the incoming electron,  $\hbar=c=1$.  In Refs.~\cite{Racah36,Racah37} the total  cross section was obtained without restrictions needed for applicability of the    Weizs\"acker-Williams approximation. However the effect of Fermi statistics for two outgoing electrons were not taken into account at that time. In Ref.~\cite{BKW} the approximate result for the  total cross section, which is in good agreement with that given by Racah~\cite{Racah36,Racah37}, was obtained. In Refs.~\cite{MU,MUT} it is shown that  the Bhabha's formula  for the total electroproduction cross section  has a good accuracy at  $\varepsilon\gtrsim 10\,$GeV. The first numerical  evaluation of  the  electroproduction cross section was performed in Ref.~\cite{Johnson}, where the differential cross section of high-energy electroproduction was obtained taking into  account  the effects of Fermi statistics. The differential cross section of high-energy electroproduction for massless leptons  was derived in \cite{Brodsky,BjCh,Henry,Homma}.

\begin{figure}
\centering
\includegraphics[width=0.8\linewidth]{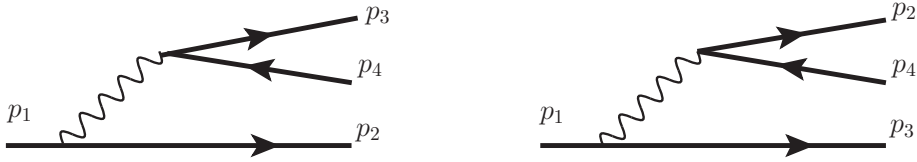}
\caption{Diagrams for the amplitude of the process $e^-Z\to e^- e^+e^-Z$. Wavy line denotes the photon propagator, straight lines denote the wave functions in the atomic field.}
\label{fig:diagrams}
\end{figure}

In all papers mentioned above the cross sections were obtained in the leading  in the parameter $\eta=Z\alpha$ approximation (in the Born approximation),  where $Z$ is the atomic charge number and $\alpha$ is the fine-structure constant. The Coulomb corrections for the differential cross section of electroproduction (the difference between the exact in $\eta$ result and the Born result) have not been derived till now. However, it is well known from the results for  the differential cross sections of photoproduction and bremsstrahlung that the Coulomb corrections may drastically change the result for heavy   atoms \cite{BM1954,DBM1954,OlsenMW1957}. It is very difficult to calculate the Coulomb corrections to the electroproduction cross section because the amplitude of  this process contains four wave functions in the atomic field, see Fig.~\ref{fig:diagrams} where the corresponding Feynman diagrams in the Furry representation are shown.
Even in the case of photoproduction and bremsstrahlung, where the corresponding amplitudes contain only two wave functions in the atomic field, exact calculations of the  Coulomb corrections for any energies are very complicated task. Fortunately, the use of the quasiclassical approximation for the electron  wave  and  Green's functions in the atomic field significantly  simplifies calculations at high electron energies (though does not make them simple).

At high  energies and small angles between outgoing and incoming particles, the main contribution to the processes in the atomic field is given by  large angular momenta of the particles.
The quasiclassical approximation provides a possibility to account for the contribution of this  large angular-momenta region. For the Coulomb potential, the wave functions  in the leading quasiclassical approximation  are the famous Furry-Sommerfeld-Maue wave functions \cite{Fu, ZM} (see also Ref.~\cite{BLP1982}). For the atomic potential, the  wave functions and the Green's functions in the leading and next-to-leading quasiclassical approximation have been derived in Refs.~\cite{LMS00, KM2015}. Using the  quasiclassical approximation the exact in $\eta$ differential cross sections for photoproduction and bremsstrahlung  in the atomic field have been obtained in Refs.~\cite{BM1954,DBM1954,OlsenMW1957,LMSS2005} in the leading quasiclassical approximation. In Refs.~\cite{KM2015} and \cite{LMS2012} the differential cross sections for bremsstrahlung and  photoproduction have been obtained exactly in $\eta$  with the first quasiclassical corrections taken into account. In Refs.~\cite{KLM2014} and \cite{KLM2015d} the cross section of $e^+e^-$ photoproduction  accompanied by bremsstrahlung  and the cross section of double bremsstrahlung have been obtained exactly in $\eta$ in the leading quasiclassical approximation.

In the present paper, we apply the quasiclassical approach to investigate, exactly in $\eta$, the differential cross section of high-energy electroproduction. The Coulomb corrections to the cross section of photoproduction are determined by the region of small impact parameters $\rho\sim \lambda_C=1/m$, while the Coulomb corrections to the cross section of bremsstrahlung are determined by large impact parameters $\rho\sim \min\{\lambda_C \varepsilon\varepsilon'/(m\omega),\,r_{scr}\}$, where  $\omega$ is the energy of emitted photon,  $\varepsilon'=\varepsilon-\omega$, and $r_{scr}\sim \lambda_C Z^{-1/3}/\alpha$ is the screening radius.  For the differential cross section of electroproduction, both regions of small and large impact parameters give the contribution to the Coulomb corrections.
We show that  the Coulomb corrections for heavy atoms drastically change the result compared with that obtained in the Born approximation.

\section{General discussion}\label{general}

The differential cross section of high-energy electroproduction  in an atomic  field   reads~\cite{BLP1982}
\begin{equation}\label{eq:cs}
d\sigma=\frac{\alpha^2}{(2\pi)^8}\varepsilon_2^2\varepsilon_3^2\varepsilon_4^2\,d\varepsilon_3d\varepsilon_4\,d\Omega_{2}\,d\Omega_{3}d\Omega_{4}\,|{\cal T}|^{2}\,,
\end{equation}
where $d\Omega_{2}$, $d\Omega_{3}$ are  the solid angles corresponding to the momenta   $\bm p_2$ and $\bm p_3$ of the outgoing electrons,  $d\Omega_{4}$ is  the solid angle corresponding to the positron momentum $\bm p_4$,  $\bm p_1$ is the incoming electron momentum (see Fig.\ref{fig:diagrams}), $\varepsilon_{1}=\varepsilon_{2}+\varepsilon_{3}+\varepsilon_{4}$ is the incoming electron energy, and $\varepsilon_{i}=\sqrt{{p}_i^2+m^2}$. Below  we assume that $\varepsilon_{i}\gg m$.   The matrix element ${\cal T}$ reads
\begin{align}\label{M12}
&{\cal T}=T+\widetilde{T}\,,\quad \widetilde{T}=-T(2\leftrightarrow 3)\, ,\nonumber\\
&T=\sum_{a,b=1}^3\int \frac{d\bm k}{(2\pi)^3}{\cal D}^{ab}j^a\,J^b\,,\quad
{\cal D}^{ab}=-\frac{4\pi}{\omega^2-k^2+i0}\left(\delta^{ab}-\frac{k^ak^b}{\omega^2}\right)\,,\nonumber\\
&\bm j=\int d\bm r\, e^{-i\bm k\cdot\bm r}\,\bar u_{\bm p_2 }^{(-)}(\bm r )\bm\gamma\,u _{\bm p_1}^{(+)}(\bm r )\,,\quad
\bm J=\int d\bm r\, e^{i\bm k\cdot\bm r}\,\bar u_{\bm p_3 }^{(-)}(\bm r )\bm\gamma\,v _{\bm p_4}^{(+)}(\bm r )\,,\nonumber\\
&\omega=\varepsilon_{1}-\varepsilon_{2}=\varepsilon_{3}+\varepsilon_{4}\,,
\end{align}
where $D^{\mu\nu}$ is a photon propagator ($D^{\mu0}=0$), $\gamma^\nu$ are the Dirac matrices, $ u_{\bm p}^{(+)}(\bm r )$  and $u_{\bm p}^{(-)}(\bm r )$ are the positive-energy  solutions of the Dirac equation in the  atomic potential  $V(r)$,
$ v_{\bm p}^{(+)}(\bm r )$ is the negative-energy  solution of the Dirac equation in the  atomic potential,  the superscripts $(-)$ and $(+)$ indicate that the asymptotic forms of  the wave functions contain at large distances  $ r$, in addition to the plane wave, the spherical convergent and divergent waves, respectively.
We  calculate the matrix element  of  electroproduction in the leading  quasiclassical approximation. In this case the wave functions have the
form \cite{LMS00}
\begin{align}\label{wfD1}
&\bar u_{\bm p }^{(-)}(\bm r )=\bar u_{\bm p }[f_0(\bm r,\bm p)-\bm\alpha\cdot\bm f_1(\bm r,\bm p)]\,,\nonumber\\
&u _{\bm p}^{(+)}(\bm r )=[g_0(\bm r,\bm p)-\bm\alpha\cdot\bm g_1(\bm r,\bm p)]u _{\bm p}\,,\nonumber\\
&v _{\bm p}^{(+)}(\bm r )=[G_0(\bm r,\bm p)+\bm\alpha\cdot\bm G_1(\bm r,\bm p)]v _{\bm p}\,,\nonumber\\
& u_{\bm p}=\sqrt{\frac{\varepsilon_p+m}{2\varepsilon_p}}
\begin{pmatrix}
\phi\\
\dfrac{{\bm \sigma}\cdot {\bm
p}}{\varepsilon_p+m}\phi
\end{pmatrix}\,,\quad
v_{\bm p}=\sqrt{\frac{\varepsilon_p+m}{2\varepsilon_p}}
 \begin{pmatrix}\dfrac{{\bm \sigma}\cdot {\bm p}}{\varepsilon_p+m}\chi\\
\chi
\end{pmatrix}\,,
\end{align}
where $\phi$ and  $\chi$  are spinors, $\bm\alpha=\gamma^0\bm\gamma$, and $\bm\sigma$ are the Pauli matrices.
The functions $f_0$ and  $\bm f_{1}$   read
\begin{align}\label{fgqc}
& f_0(\bm r,\bm p)=-\frac{i}{\pi}e^{-i\bm p\cdot\bm r}\int d\bm Q \exp\left[iQ^2-i
\int_0^\infty dx V(\bm r_p) \right]\,,\nonumber\\
&\bm f_1(\bm r,\bm p)=\frac{1}{2\varepsilon_p}(i\bm\nabla-\bm p)f_0(\bm r,\bm p)\,,\nonumber\\
&\bm r_p= \bm r+x\bm n_{ p}+\sqrt{\frac{2x}{\varepsilon_p}}\,\bm Q \,,
\end{align}
where $\bm Q$ is a two-dimensional vector perpendicular to the vector $\bm n_p=\bm p/p$.  The expressions for the  functions   $g_0$ and $\bm g_{1}$ follow from the relations
\begin{equation}
g_0(\bm r,\bm p)=f_0(\bm r,-\bm p)\,,\quad \bm g_1(\bm r,\bm p)= f_1(\bm r,-\bm p)\,,
\end{equation}
and the expressions for the functions  $G_0$ and $\bm G_{1}$ can be obtained from the functions $f_0$ and  $\bm f_{1}$, respectively, by the replacement $V(\bm r_p)\rightarrow -V(\bm r_p)$.

It is convenient to calculate the matrix element for definite helicities of the particles. Let $\mu_i$  be a  sign of the helicity of a particle with the momentum $\bm p_i$.   We  direct   the $z$-axis along a unit vector $\bm\nu$ assuming that the angles between $\bm\nu$ and $\bm p_i$ are small. The final result will be independent of the direction of $\bm\nu$. Then,  to calculate the matrix element, we use the matrices
${\cal F}=u_{{\bm p_1}\,\mu_1}\bar{u}_{{\bm p_2}\,\mu_2}$ and  $\widetilde{\cal F}=v_{{\bm p}_4\,\mu_4}\bar{u}_{{\bm p_3}\,\mu_3}$ \cite{KLM2014, KM2015}
\begin{align}\label{calf}
&{\cal F}=\frac{1}{8}(a_{\mu_1\mu_2}+\bm\Sigma\cdot \bm b_{\mu_1\mu_2})[\gamma^0(1+P_1P_2)+\gamma^0\gamma^5(P_1+P_2)+(1-P_1P_2)-\gamma^5(P_1-P_2)],\nonumber\\
&\widetilde{\cal F}=\frac{1}{8}(\widetilde a_{\mu_3\mu_4}+\bm\Sigma\cdot \widetilde{ \bm b}_{\mu_3\mu_4})[\gamma^0(P_3-P_4)+\gamma^0\gamma^5(1-P_3P_4)-(P_3+P_4)-\gamma^5(1+P_3P_4)].
\end{align}
Here $P_i=\mu_ip_i/(\varepsilon_i+m)$,  $\bm\Sigma=-\gamma^5\bm\alpha$, $\gamma^5=-i\gamma^0\gamma^1\gamma^2\gamma^3$;
$a_{\mu_1\mu_2}$, $\bm b_{\mu_1\mu_2}$, $\widetilde a_{\mu_3\mu_4}$, and $\widetilde{ \bm b}_{\mu_3\mu_4}$ are
\begin{align}\label{ab}
&a_{\mu\mu}=1\,,\quad  a_{\mu\bar\mu}=\frac{\mu}{\sqrt{2}}\bm s_\mu\cdot\bm\theta_{12}\,,\nonumber\\
&\bm b_{\mu\mu}=\mu\bm\nu+\frac{\mu}{2}(\bm\theta_{1}+\bm\theta_{2})+\frac{i}{2}[\bm\theta_{12}\times\bm\nu]\,,\nonumber\\
&\bm b_{\mu\bar\mu}=\sqrt{2}\bm s_\mu-\frac{1}{\sqrt{2}}(\bm s_\mu,\bm\theta_{1}+\bm\theta_{2})\bm\nu\,,\nonumber\\
&\widetilde a_{\mu\bar\mu}=\mu\,,\quad\widetilde a_{\mu\mu}=-\frac{1}{\sqrt{2}}\bm s_\mu^*\cdot\bm\theta_{34}\,,\nonumber\\
&\widetilde {\bm b}_{\mu\bar\mu}=\bm\nu+\frac{1}{2}(\bm\theta_{3}+\bm\theta_{4})-\frac{i\mu}{2}[\bm\theta_{34}\times\bm\nu]\,,\nonumber\\
&\widetilde {\bm b}_{\mu\mu}=-\mu\sqrt{2}\bm s_\mu^*+\frac{\mu}{\sqrt{2}}(\bm s_\mu^*,\bm\theta_{3}+\bm\theta_{4})\bm\nu\,,\nonumber\\
& \bm s_{\mu}=\frac{1}{\sqrt{2}}(\bm e_x+i\mu\bm e_y)\,,
\end{align}
where  $\bar\mu=-\mu$, $\bm e_x$ and  $\bm e_y$ are two orthogonal unit vectors perpendicular to $\bm\nu$, $\bm \theta_i=\bm p_{i\perp}/p_i$, $\bm \theta_{ij}=\bm \theta_i-\bm \theta_j$, and
the notation $\bm X_\perp= \bm X-(\bm X\cdot\bm\nu)\bm\nu$ for any vector $\bm X$ is used.

It is convenient to write the  photon propagator  ${\cal D}^{ab}$ as follows:
\begin{align}\label{Dnew}
&{\cal D}^{ab}={\cal D}^{ab}_\perp+{\cal D}^{ab}_\parallel\,,\nonumber\\
&{\cal D}^{ab}_\perp=-\frac{4\pi}{\omega^2-k^2+i0}\left(\delta^{ab}-\frac{k^ak^b}{k^2}\right)=-\frac{4\pi}{\omega^2-k^2+i0}\sum_{\lambda=\pm}s_\lambda^{a*}s_\lambda^{b}\,,\nonumber\\
&{\cal D}^{ab}_\parallel=-\frac{4\pi}{\omega^2k^2}k^ak^b=-\frac{4\pi}{\omega^2}\nu^a\nu^b\,,
\end{align}
where we direct the vector $\bm\nu$ along the vector $\bm k$. Substituting this expression to Eq.\eqref{M12} we obtain for $T$
\begin{align}\label{M1new}
&T=T_\perp+T_\parallel\,,\nonumber\\
&T_\perp=-4\pi\sum_{\lambda=\pm}\int \frac{d\bm k\,j_\lambda\,J_\lambda}{(2\pi)^3(\omega^2-k^2+i0)}\,,\nonumber\\
&T_\parallel=-\frac{4\pi}{\omega^2}\int \frac{d\bm k}{(2\pi)^3}j_\parallel\,J_\parallel\,,\nonumber\\
&j_\lambda=\bm j\cdot\bm s_\lambda^*\,,\quad J_\lambda=\bm J\cdot\bm s_\lambda\,,\quad j_\parallel=\bm j\cdot\bm\nu\,,\quad J_\parallel=\bm J\cdot\bm\nu\,.
\end{align}
The functions  $\bm j$ and $\bm J$ correspond to  the matrix elements of virtual photon bremsstrahlung and pair production by virtual photon, respectively. The calculation of these functions can be performed in the same way as it was done for the  real bremsstrahlung cross section \cite{LMSS2005, KM2015} and for the  pair production cross section by a real photon~\cite{KLM2014}.
As a result we obtain for the matrix elements $j_\lambda$ and $j_\parallel$ of virtual bremsstrahlung
\begin{align}\label{jlambda}
&j_\lambda=-A(\bm\Delta)\Bigg[\delta_{\mu_1\mu_2}(\varepsilon_1\delta_{\lambda\mu_1}+\varepsilon_2\delta_{\lambda\bar\mu_1})
\left(\bm s_\lambda^*, \frac{\bm\theta_2}{\varepsilon_1 D_1}+ \frac{\bm\theta_1}{\varepsilon_2 D_2}\right)\,\nonumber\\
&+\delta_{\mu_1\bar\mu_2}\delta_{\lambda\mu_1}\frac{m\omega\mu_1}{\sqrt{2}\varepsilon_1\varepsilon_2}\left(\frac{1}{D_1}+\frac{1}{D_2}\right)\Bigg]\,,\nonumber\\
&j_\parallel=-A(\bm\Delta)\delta_{\mu_1\mu_2}\left(\frac{1}{D_1}+\frac{1}{D_2}\right)\,,\nonumber\\
&A(\bm\Delta)=-\frac{i}{\Delta_{\perp}^2}\int d\bm r\,\exp[-i\bm\Delta\cdot\bm r-i\chi(\rho)]\bm\Delta_{\perp}\cdot\bm\nabla_\perp V(r)\,,\nonumber\\
&\chi(\rho)=\int_{-\infty}^\infty dz\,V(\sqrt{z^2+\rho^2})\,,\nonumber\\
&D_1=\frac{\Delta_{\perp}^2}{2\varepsilon_1}+\bm n_1\cdot\bm\Delta-i0\,,\quad D_2=\frac{\Delta_\perp^2}{2\varepsilon_2}-\bm n_2\cdot\bm\Delta-i0\,, \nonumber\\
&\bm\Delta=\bm k+\bm p_2-\bm p_1\,,\quad  \bm n_i=\bm p_i/p_i\,.
\end{align}
At $\Delta_\perp\gg\max(\Delta_\parallel,r_{scr}^{-1})$ , where $\Delta_\parallel=\bm\Delta\cdot\bm\nu$ and $r_{scr}$ is a screening radius, the function $A(\bm \Delta)$ is independent of the potential shape (see Ref. \cite{LMSS2005}). It has the following asymptotic form
\begin{align}\label{ACJCas}
&A_{as}(\bm\Delta)=-\frac{4\pi\eta (L\Delta_\perp)^{2i\eta}\Gamma(1-i\eta)}{\Delta_\perp^2\Gamma(1+i\eta)}\,,
\end{align}
where  $\Gamma(x)$ is the Euler $\Gamma$ function, a specific value of $L\sim \max(\Delta_\parallel,r_{scr}^{-1})$ is irrelevant because the factor $L^{2i\eta}$ disappears in $|T_{tot}|^2$.
At $\Delta_\perp\lesssim\max(\Delta_\parallel,r_{scr}^{-1})$ ,  the function $A(\bm \Delta)$ strongly depends on  $\Delta_\parallel$ and the shape of the atomic potential  \cite{LMSS2005}.

The matrix elements $J_\lambda$ and $J_\parallel$ of pair production by virtual photon read
\begin{align}\label{Jlambda}
&J_\lambda=J_\lambda^{(0)}+J_\lambda^{(1)}\,,\quad  J_\parallel=J_\parallel^{(0)}+J_\parallel^{(1)}\,, \nonumber\\
&J_\lambda^{(0)}=(2\pi)^3\delta(\bm p_3+\bm p_4-\bm k)\left[\delta_{\mu_3\bar\mu_4}\left(\bm s_\lambda, \delta_{\lambda\mu_3}\bm\theta_4+\delta_{\lambda\mu_4}\bm\theta_3\right)
-\delta_{\mu_3\mu_4}\delta_{\lambda\mu_3}\frac{m\omega\mu_3}{\sqrt{2}\varepsilon_3\varepsilon_4}\right]\,,\nonumber\\
&J_\lambda^{(1)}=\frac{i\varepsilon_3\varepsilon_4}{2\pi\omega}\int\limits_0^\infty \frac{dz}{z}e^{iz(p_3+p_4-k+i 0)}\iint\,d^2Q_3d^2Q_4\,{\cal J}e^{i\Phi} \nonumber\\
&\times\left[\frac{\delta_{\mu_3\bar\mu_4}}{\omega z}(\bm s_\lambda\cdot \bm Q_{34})(\varepsilon_3\delta_{\lambda\mu_3}-\varepsilon_4\delta_{\lambda\mu_4})+
\delta_{\mu_3\mu_4}\delta_{\lambda\mu_3}\frac{m\omega\mu_3}{\sqrt{2}\varepsilon_3\varepsilon_4}\right]\,,\nonumber\\
&J_\parallel^{(0)}=(2\pi)^3\delta(\bm p_3+\bm p_4-\bm k)\delta_{\mu_3\bar\mu_4}\,,\nonumber\\
&J_\parallel^{(1)}=-\frac{i\varepsilon_3\varepsilon_4}{2\pi\omega}\int\limits_0^\infty \frac{dz}{z}e^{iz(p_3+p_4-k+i0)}\iint\,d^2Q_3d^2Q_4\,{\cal J} e^{i\Phi} \delta_{\mu_3\bar\mu_4}\,,\nonumber\\
&{\cal J}=e^{i[\chi(Q_4)-\chi(Q_3)]}-1\,,\quad \Phi=\frac{\varepsilon_3\varepsilon_4}{2\omega z} Q_{34}^2-
(\varepsilon_3\bm Q_3\cdot\bm\theta_3+ \varepsilon_4\bm Q_4\cdot\bm\theta_4)\,,
\end{align}
where $\bm Q_{34}=\bm Q_{3}-\bm Q_{4}$. The matrix elements  $J_\lambda^{(0)}$ and $J_\parallel^{(0)}$ correspond to the virtual photon decay into $e^+e^-$ pair  noninteracting with the atomic field, while the matrix elements  $J_\lambda^{(1)}$ and $J_\parallel^{(1)}$ correspond to production of   pair interacting with the atomic field.
Then we substitute Eqs.~\eqref{jlambda} and \eqref{Jlambda} in Eq.~\eqref{M12} and write the amplitudes $T_\perp$ and $T_\parallel$ as follows
\begin{align}\label{T01}
&T_\perp=T_\perp^{(0)}+T_\perp^{(1)}\,,\quad T_\parallel=T_\parallel^{(0)}+T_\parallel^{(1)}\,.
\end{align}
Integrating  over $\bm k$, we obtain for the  terms $T_\perp^{(0)}$ and $T_\parallel^{(0)}$:
\begin{align}\label{T0}
&T_\perp^{(0)}=\frac{8\pi\varepsilon_3\varepsilon_4A(\bm\Delta_0)}{m^2\omega^2+\varepsilon_3^2\varepsilon_4^2\theta_{34}^2}\Big\{\delta_{\mu_1\mu_2}\delta_{\mu_3\bar\mu_4}
\Big[\frac{\varepsilon_3}{\omega^2}(\bm s_{\mu_3}^*\cdot \bm X)(\bm s_{\mu_3}\cdot\bm\theta_{34})(\varepsilon_1\delta_{\mu_1\mu_3}+\varepsilon_2\delta_{\mu_1\mu_4})\nonumber\\
&-\frac{\varepsilon_4}{\omega^2}(\bm s_{\mu_4}^*\cdot \bm X)(\bm s_{\mu_4}\cdot\bm\theta_{34}) (\varepsilon_1\delta_{\mu_1\mu_4}+\varepsilon_2\delta_{\mu_1\mu_3})\Big]\nonumber\\
&-\frac{m\mu_1}{\sqrt{2}\varepsilon_1\varepsilon_2}R\delta_{\mu_1\bar\mu_2}\delta_{\mu_3\bar\mu_4}
(\bm s_{\mu_1}\cdot\bm\theta_{34})(-\varepsilon_3\delta_{\mu_1\mu_3}+\varepsilon_4\delta_{\mu_1\mu_4})\nonumber\\
&+\frac{m\mu_3}{\sqrt{2}\varepsilon_3\varepsilon_4}\delta_{\mu_1\mu_2}\delta_{\mu_3\mu_4}(\bm s_{\mu_3}^*\cdot\bm X)(\varepsilon_1\delta_{\mu_3\mu_1}+\varepsilon_2\delta_{\mu_3\bar\mu_1})
+\frac{m^2\omega^2}{2\varepsilon_1\varepsilon_2\varepsilon_3\varepsilon_4}R\delta_{\mu_1\bar\mu_2}\delta_{\mu_3\mu_4}\delta_{\mu_1\mu_3}\Big\}\,,\nonumber\\
&T_\parallel^{(0)}=-\frac{8\pi }{\omega^2}A(\bm\Delta_0)R\delta_{\mu_1\mu_2}\delta_{\mu_3\bar\mu_4}\,.
\end{align}
Here
\begin{align}\label{T0not}
& \bm\Delta_0=\bm p_2+\bm p_3+\bm p_4-\bm p_1\,,\quad \bm\Delta_{0\perp}=\varepsilon_2\bm\theta_{21}+\varepsilon_3\bm\theta_{31}+\varepsilon_4\bm\theta_{41}\,,\nonumber\\
&\Delta_{0\parallel}=-\frac{1}{2}\left[m^2\omega\left(\frac{1}{\varepsilon_1\varepsilon_2}+\frac{1}{\varepsilon_3\varepsilon_4}\right)+\varepsilon_2\theta_{21}^2+
\varepsilon_3\theta_{31}^2+\varepsilon_4\theta_{41}^2\right]\,,\nonumber\\
&R=\frac{1}{d_1d_2}[\Delta^2_{0\perp} (\varepsilon_1+\varepsilon_2)+2\varepsilon_1\varepsilon_2(\bm\theta_{12}\cdot\bm\Delta_{0\perp})]\,,\nonumber\\
&\bm X=\frac{1}{d_1}(\varepsilon_3\bm\theta_{23}+\varepsilon_4\bm\theta_{24})-\frac{1}{d_2}(\varepsilon_3\bm\theta_{13}+\varepsilon_4\bm\theta_{14})\,,\nonumber\\
&d_1=m^2\omega\varepsilon_1\left(\frac{1}{\varepsilon_1\varepsilon_2}+\frac{1}{\varepsilon_3\varepsilon_4}\right)+\varepsilon_2\varepsilon_3\theta_{23}^2
+\varepsilon_2\varepsilon_4\theta_{24}^2+\varepsilon_3\varepsilon_4\theta_{34}^2\,,\nonumber\\
&d_2=m^2\omega\varepsilon_2\left(\frac{1}{\varepsilon_1\varepsilon_2}+\frac{1}{\varepsilon_3\varepsilon_4}\right)+\varepsilon_2\varepsilon_3\theta_{31}^2
+\varepsilon_2\varepsilon_4\theta_{41}^2+(\varepsilon_3\bm\theta_{31}+\varepsilon_4\bm\theta_{41})^2\,.
\end{align}
These amplitudes correspond to  production of $e^+e^-$ pair non-interacting with the atomic field, so that  they  have  the  dependence on the atomic potential similar to that of the  bremsstrahlung amplitude, see, e.g.,\cite{LMSS2005,KLM2015d}.

To derive the terms $T_\perp^{(1)}$ and $T_\parallel^{(1)}$, we take the   integral over $k_z$ by
 closing  the contour  of integration in the lower half-plane of the complex variable $k_z$. Then  the main contribution to the integral is given by the pole of the function $1/D_2$ in Eq.~\eqref{jlambda}. We have
\begin{align}\label{T1}
&T_\perp^{(1)}=-\frac{\varepsilon_1\varepsilon_3\varepsilon_4}{2\pi^2\omega} \int\frac{d\bm\Delta_\perp\, A(\bm\Delta_\perp)}{m^2\omega^2+\varepsilon_1^2Y^2}
\int\limits_0^\infty \frac{dz}{z}\exp\Big(-\frac{iz}{2}\Phi_1\Big)\iint\,d^2Q_3d^2Q_4\,{\cal J} e^{i\Phi_2} \nonumber\\
&\times \Big\{\frac{\delta_{\mu_1\mu_2}\delta_{\mu_3\bar\mu_4}}{\omega^2z} \big[ \varepsilon_1(\varepsilon_3 \delta_{\mu_1\mu_3}-\varepsilon_4 \delta_{\mu_1\mu_4})
(\bm s_{\mu_1}^*\cdot\bm Y)(\bm s_{\mu_1}\cdot\bm Q_{34})\,\nonumber\\
&+\varepsilon_2(\varepsilon_3 \delta_{\mu_1\bar\mu_3}-\varepsilon_4 \delta_{\mu_1\bar\mu_4})(\bm s_{\mu_1}\cdot\bm Y)(\bm s_{\mu_1}^*\cdot\bm Q_{34})  \big]-\delta_{\mu_1\bar\mu_2}\delta_{\mu_3\bar\mu_4}\frac{m\mu_1}{\sqrt{2}\varepsilon_1 z}(\varepsilon_3 \delta_{\mu_1\mu_3}-\varepsilon_4 \delta_{\mu_1\mu_4})(\bm s_{\mu_1}
\cdot\bm Q_{34})\nonumber\\
&+\delta_{\mu_1\mu_2}\delta_{\mu_3\mu_4}\frac{m\mu_3}{\sqrt{2}\varepsilon_3\varepsilon_4}(\varepsilon_1 \delta_{\mu_1\mu_3}+\varepsilon_2 \delta_{\mu_1\bar\mu_3})(\bm s_{\mu_3}^*\cdot\bm Y)
-\frac{m^2\omega^2}{2\varepsilon_1\varepsilon_3\varepsilon_4}\delta_{\mu_1\bar\mu_2}\delta_{\mu_3\mu_4}\delta_{\mu_1\mu_3}\Big\}\,,\nonumber\\
&T_\parallel^{(1)}=\frac{\varepsilon_3\varepsilon_4}{2\pi^2\omega^3} \int d\bm\Delta_\perp\, A(\bm\Delta_\perp)
\int\limits_0^\infty \frac{dz}{z}\exp\Big(-\frac{iz}{2}\Phi_1\Big)\iint\,d^2Q_3d^2Q_4\,{\cal J} e^{i\Phi_2}\delta_{\mu_1\mu_2}\delta_{\mu_3\bar\mu_4}\,,\nonumber\\
&\Phi_1=m^2\omega\Big(\frac{1}{\varepsilon_1\varepsilon_2}+\frac{1}{\varepsilon_3\varepsilon_4}\Big)
+\frac{\varepsilon_1}{\varepsilon_2\omega}Y^2\,,\quad \bm Y=\bm\Delta_\perp-\varepsilon_2\bm\theta_{21}\,,\nonumber\\
&\Phi_2=\frac{\varepsilon_3\varepsilon_4}{2\omega z} Q_{34}^2+
\frac{\varepsilon_3}{\omega}\bm Q_3\cdot(\bm Y-\omega\bm\theta_{31})+ \frac{\varepsilon_4}{\omega}\bm Q_4\cdot(\bm Y-\omega\bm\theta_{41})\,.
\end{align}
Here $\cal J$ is given in Eq.~\eqref{Jlambda},  $A(\bm\Delta_\perp)$ is the function $A(\bm\Delta)$ at $\Delta_\parallel=0$, and the integration over $\bm\Delta_\perp$ is the integration over two-dimensional vector perpendicular to $z$-axis.

\section{Coulomb field}\label{CF}
Let us consider the region $\min(\Delta_0,\,\Delta_1)\gg  r_{scr}^{-1}$, where $r_{scr}$ is the screening radius and $\Delta_1=\sqrt{(\varepsilon_2\theta_{21})^2+(m\omega/\varepsilon_1)^2}$. In this case we can neglect the effect of screening and use  the Coulomb potential  $V_C(r)=-\eta/r$ instead  of the atomic potential $V(r)$. Then we have
\begin{align}\label{ACJC}
&\chi(\rho)=-2\eta\ln(2L/\rho)\,,\quad{\cal J}=\left(\frac{Q_4}{Q_3}\right)^{2i\eta}-1 \,,\nonumber\\
&A_C(\bm \Delta)=-\frac{4\pi\eta (L\Delta)^{2i\eta}}{\Delta^2}\Gamma(1-i\eta)\Gamma(2-i\eta)F\left(1-i\eta, i\eta,2,\frac{\Delta_\perp^2}{\Delta^2}\right)
\end{align}
where  $F(a,b,c,x)$ is the hypergeometric function, $\Gamma(x)$ is the Euler $\Gamma$ function, $L\sim  r_{scr}$ and $A_C(\bm \Delta)$ is $A(\bm \Delta)$ in (\ref{jlambda}) for the Coulomb potential.
At $\Delta_\perp\gg\Delta_\parallel$ the function  $A_C(\bm \Delta)$ coincides with $A_{as}(\bm \Delta)$, Eq.~\eqref{ACJCas}.

In the Coulomb field the amplitudes $T_\perp^{(0)}$ and $T_\parallel^{(0)}$ are given by Eq.~\eqref{T0} with the replacement $A(\bm \Delta_0)\rightarrow A_C(\bm \Delta_0)$.
To derive the amplitudes  $T_\perp^{(1)}$ and $T_\parallel^{(1)}$ from Eq.~\eqref{T1}, we integrate  over the variables $\bm Q_3+\bm Q_4$  and $z$  in the same way as it was done in the case of the photoproduction cross section (see, e.g., \cite{KLM2014}). The integration over $\bm Q_{34}=\bm Q_3-\bm Q_4  $  is  performed with the use of the Feynman parametrization. As a result we obtain
\begin{align}\label{T1C}
&T_\perp^{(1)}=\frac{8i\eta\varepsilon_1}{\omega}|\Gamma(1-i\eta)|^2 \int\frac{d\bm\Delta_\perp\, A_{as}(\bm\Delta_\perp)}{Q^2 M^2\,(m^2\omega^2+\varepsilon_1^2Y^2)}\left(\frac{\xi_2}{\xi_1}\right)^{i\eta}
{\cal M}\,, \nonumber\\
&{\cal M}=-\frac{\delta_{\mu_1\mu_2}\delta_{\mu_3\bar\mu_4}}{\omega} \big[ \varepsilon_1(\varepsilon_3 \delta_{\mu_1\mu_3}-\varepsilon_4 \delta_{\mu_1\mu_4})
(\bm s_{\mu_1}^*\cdot\bm Y)(\bm s_{\mu_1}\cdot\bm I_1)\,\nonumber\\
&+\varepsilon_2(\varepsilon_3 \delta_{\mu_1\bar\mu_3}-\varepsilon_4 \delta_{\mu_1\bar\mu_4})(\bm s_{\mu_1}\cdot\bm Y)(\bm s_{\mu_1}^*\cdot\bm I_1)  \big]+\delta_{\mu_1\bar\mu_2}\delta_{\mu_3\bar\mu_4}\frac{m\omega\mu_1}{\sqrt{2}\varepsilon_1 }(\varepsilon_3 \delta_{\mu_1\mu_3}-\varepsilon_4 \delta_{\mu_1\mu_4})(\bm s_{\mu_1}
\cdot\bm I_1)\nonumber\\
&+\delta_{\mu_1\mu_2}\delta_{\mu_3\mu_4}\frac{m\mu_3}{\sqrt{2}}(\varepsilon_1 \delta_{\mu_1\mu_3}+\varepsilon_2 \delta_{\mu_1\bar\mu_3})(\bm s_{\mu_3}^*\cdot\bm Y)I_0
-\frac{m^2\omega^2}{2\varepsilon_1}\delta_{\mu_1\bar\mu_2}\delta_{\mu_3\mu_4}\delta_{\mu_1\mu_3}I_0\,,\nonumber\\
&T_\parallel^{(1)}=-\frac{8i\eta\varepsilon_3\varepsilon_4}{\omega^3}|\Gamma(1-i\eta)|^2 \int \frac{d\bm\Delta_\perp\, A_{as}(\bm\Delta_\perp)}{Q^2 M^2}\left(\frac{\xi_2}{\xi_1}\right)^{i\eta}\,I_0
\delta_{\mu_1\mu_2}\delta_{\mu_3\bar\mu_4}\,,
\end{align}
where the function  $A_{as}(\bm\Delta_\perp)$ is given in Eq.~\eqref{ACJCas} and  the following notations are used
\begin{align}\label{T1Cnot}
&M^2=m^2\Big(1+\frac{\varepsilon_3\varepsilon_4}{\varepsilon_1\varepsilon_2}\Big)
+\frac{\varepsilon_1\varepsilon_3\varepsilon_4}{\varepsilon_2\omega^2}Y^2\,,\quad \bm Y=\bm\Delta_\perp-\varepsilon_2\bm\theta_{21}\,,\quad
 \bm\zeta=\frac{\varepsilon_3\varepsilon_4}{\omega}\bm\theta_{34}\,               \nonumber\\
&\bm Q=\bm\Delta_\perp-\bm\Delta_0\,,\quad
\bm q_1=\frac{\varepsilon_3}{\omega}\bm Q_\perp- \bm\zeta\,,\quad \bm q_2=
 \frac{\varepsilon_4}{\omega}\bm Q_\perp+ \bm\zeta\,,\nonumber\\
&I_0=(\xi_1-\xi_2)F(x)+(\xi_1+\xi_2-1)(1-x)\frac{F'(x)}{i\eta}\,,\nonumber\\
&\bm I_1=(\xi_1\bm q_1+\xi_2\bm q_2)F(x)+(\xi_1\bm q_1-\xi_2\bm q_2)(1-x)\frac{F'(x)}{i\eta}\,,\nonumber\\
&\xi_1=\frac{M^2}{M^2+q_1^2}\,,\quad \xi_2=\frac{M^2}{M^2+q_2^2}\,,\quad x=1-\frac{Q_\perp^2\xi_1\xi_2}{M^2}\,,\nonumber\\
&F(x)=F(i\eta,-i\eta, 1,x)\,,\quad F'(x)=\frac{\partial}{\partial x}F(x)\,.
\end{align}
Note that $Q_\parallel=-\Delta_{0\parallel}$. In contrast to the term $T^{(0)}$,  the Coulomb corrections to the term $T^{(1)}$ significantly modify the differential cross section of electroproduction not only at small $\Delta_0$ but also at $\Delta_0\sim m$.

\subsection{Born amplitude}
In the leading Born approximation, the terms  $T_{B\perp}^{(0)}$ and $T_{B\parallel}^{(0)}$ are given by Eq.~\eqref{T0} with the replacement
$$A(\bm\Delta_0)\rightarrow A_{B}(\bm\Delta_0)= -4\pi\eta /\Delta_0^2\,.$$
To derive  the terms  $T_{B\perp}^{(1)}$ and $T_{B\parallel}^{(1)}$ from Eq.~\eqref{T1C}, we use the relation
\begin{equation}
\lim_{\eta\rightarrow 0}\,\eta\int d\bm\Delta_\perp\,\Delta_\perp^{2i\eta-2}G(\bm\Delta_\perp)=-i\pi G(0)\,,
\end{equation}
where $G(\bm\Delta_\perp)$ is some function. Then we obtain
\begin{align}\label{T1CB}
&T_{B\perp}^{(1)}=\frac{8\pi\varepsilon_1\varepsilon_2 A_{B}(\bm\Delta_0)}{\omega^2 M_B^2\,(m^2\omega^2+\varepsilon_1^2\varepsilon_2^2\theta_{21}^2)}{\cal M}_B\,,\quad
T_{B\parallel}^{(1)}=-\frac{8\pi\varepsilon_3\varepsilon_4 A_{B}(\bm\Delta_0)}{\omega^3M_B^2}I_{B0}\delta_{\mu_1\mu_2}\delta_{\mu_3\bar\mu_4}\,,\nonumber\\
&{\cal M}_B= \delta_{\mu_1\mu_2}\delta_{\mu_3\bar\mu_4} \big[ \varepsilon_1(\varepsilon_3 \delta_{\mu_1\mu_3}-\varepsilon_4 \delta_{\mu_1\mu_4})
(\bm s_{\mu_1}^*\cdot\bm\theta_{21})(\bm s_{\mu_1}\cdot\bm I_{B1})\,\nonumber\\
&+\varepsilon_2(\varepsilon_3 \delta_{\mu_1\bar\mu_3}-\varepsilon_4 \delta_{\mu_1\bar\mu_4})(\bm s_{\mu_1}\cdot\bm\theta_{21})(\bm s_{\mu_1}^*\cdot\bm I_{B1})  \big]+\delta_{\mu_1\bar\mu_2}\delta_{\mu_3\bar\mu_4}\frac{m\omega^2\mu_1}{\sqrt{2}\varepsilon_1\varepsilon_2 }(\varepsilon_3 \delta_{\mu_1\mu_3}-\varepsilon_4 \delta_{\mu_1\mu_4})(\bm s_{\mu_1}
\cdot\bm I_{B1})\nonumber\\
&-\delta_{\mu_1\mu_2}\delta_{\mu_3\mu_4}\frac{m\mu_3\omega}{\sqrt{2}}(\varepsilon_1 \delta_{\mu_1\mu_3}+\varepsilon_2 \delta_{\mu_1\bar\mu_3})(\bm s_{\mu_3}^*\cdot\bm\theta_{21})I_{B0}
-\frac{m^2\omega^3}{2\varepsilon_1\varepsilon_2}\delta_{\mu_1\bar\mu_2}\delta_{\mu_3\mu_4}\delta_{\mu_1\mu_3}I_{B0}\,,\nonumber\\
&M_B^2=m^2\Big(1+\frac{\varepsilon_3\varepsilon_4}{\varepsilon_1\varepsilon_2}\Big)
+\frac{\varepsilon_1\varepsilon_2\varepsilon_3\varepsilon_4}{\omega^2}\bm\theta_{21}^2\,,\nonumber\\
&I_{B0}=\xi_{B1}-\xi_{B2}\,,\quad \bm I_{B1}=\xi_{B1}\bm q_{B1}+\xi_{B2}\bm q_{B2}\,,\nonumber\\
&\xi_{B1}=\frac{M_B^2}{M_B^2+q_{B1}^2}\,,\quad \xi_{B2}=\frac{M_B^2}{M_B^2+q_{B2}^2}\,,\quad \bm\zeta=\frac{\varepsilon_3\varepsilon_4}{\omega}\bm\theta_{34}\,,\nonumber\\
&\bm q_{B1}=-\frac{\varepsilon_3}{\omega}\bm\Delta_{0\perp}- \bm\zeta\,,\quad \bm q_{B2}= -\frac{\varepsilon_4}{\omega}\bm\Delta_{0\perp}+ \bm\zeta\,.
\end{align}
The region, which gives    the leading logarithmic contribution to the Born cross section, is
\begin{eqnarray}\label{reg1}
 m\ll\omega\ll \varepsilon_1\,,\quad \dfrac{m^2}{\omega}\ll\Delta_{0\perp}\ll m\,,\quad \dfrac{m\omega}{\varepsilon_1}\ll \varepsilon_1\theta_{21}\ll m \,,\quad
\zeta\sim m\,.
\end{eqnarray}
In this region
\begin{align}\label{T1Cas11}
&T_{B\perp}^{(1)}=\frac{32\pi^2\eta\varepsilon_1 \delta_{\mu_1\mu_2}}{\omega^2(m^2+\zeta^2)\Delta_{1\perp}^2\Delta_{0\perp}^2}
{\cal M}_{as}\,,\nonumber\\
&{\cal M}_{as}=\delta_{\mu_3\bar\mu_4}
\Big[\varepsilon_3(\bm s_{\mu_3}^*\cdot\bm\theta_{21})(\bm s_{\mu_3}\cdot \bm X_0)   -\varepsilon_4 (\bm s_{\mu_4}^*\cdot\bm\theta_{21})(\bm s_{\mu_4}\cdot \bm X_0)
 \Big]\nonumber\\
 &-\mu_3\sqrt{2}m\omega\delta_{\mu_3\mu_4}(\bm s_{\mu_3}^*\cdot\bm\theta_{21})\frac{(\bm \zeta\cdot\bm\Delta_{0\perp})}{m^2+\zeta^2}\,,\nonumber\\
&\bm X_0=\bm\Delta_{0\perp}-\frac{2\bm \zeta (\bm \zeta\cdot\bm\Delta_{0\perp})}{m^2+\zeta^2}\,,
\end{align}
and the terms $T_{B\perp}^{(0)}$, $T_{B\parallel}^{(0)}$, and $T_{B\parallel}^{(1)}$  are suppressed   compared with $T_{B\perp}^{(1)}$.
Substituting the matrix element \eqref{T1Cas11}  in (\ref{eq:cs}) and performing  the integration over the region  \eqref{reg1} with the logarithmic accuracy,  we obtain the well-known result \cite{Bhabha2,Racah37}  for the leading logarithmic contribution to the total cross section of high-energy electroproduction
\begin{align}\label{csl3}
\sigma=\frac{28\eta^2\alpha^2}{27\pi m^2}\ln^3\dfrac{\varepsilon_1}{m}\,.
\end{align}

\subsection{Asymptotic forms  of the exact in $\eta$ amplitudes in the Coulomb field.}
In the region \eqref{reg1} we have  $|T|=|T_B|$ , where $T$ is the exact in $\eta$ asymptotic result and  $T_B$  is given by \eqref{T1Cas11}.
For large $p_\perp$ of  all outgoing particles,  $\varepsilon_2\theta_{21}\gg m$, $\varepsilon_3\theta_{31}\gg m$,  and $\varepsilon_4\theta_{41}\gg m$, but $\Delta_0\lesssim m$ we have
\begin{align}\label{T0Cas}
&T_\perp^{(0)}=\frac{8\pi A_{C}(\bm\Delta_0)}{\varepsilon_1\varepsilon_2 \varepsilon_3\varepsilon_4\omega c_1\theta_{34}^2}\delta_{\mu_1\mu_2}\delta_{\mu_3\bar\mu_4}
\Big[\varepsilon_3(\bm s_{\mu_3}^*\cdot \bm X_1)(\bm s_{\mu_3}\cdot\bm\theta_{34})(\varepsilon_1\delta_{\mu_1\mu_3}+\varepsilon_2\delta_{\mu_1\mu_4})\nonumber\\
&-\varepsilon_4(\bm s_{\mu_4}^*\cdot \bm X_1)(\bm s_{\mu_4}\cdot\bm\theta_{34}) (\varepsilon_1\delta_{\mu_1\mu_4}+\varepsilon_2\delta_{\mu_1\mu_3})\Big]\,,\nonumber\\
&T_\parallel^{(0)}=\frac{16\pi A_{C}(\bm\Delta_0)}{\omega^2c_1^2}\delta_{\mu_1\mu_2}\delta_{\mu_3\bar\mu_4}\,(\bm\theta_{21}\cdot\bm\Delta_{0\perp})\,,\nonumber\\
&c_1=\varepsilon_2\theta_{21}^2+\varepsilon_3\theta_{31}^2+\varepsilon_4\theta_{41}^2\,,\quad
\bm X_1=\bm\Delta_{0\perp}-\frac{2\varepsilon_1\varepsilon_2}{\omega c_1}(\bm\theta_{21}\cdot\bm\Delta_{0\perp})\bm\theta_{21}\,.
\end{align}
In this region the terms  $T_\perp^{(1)}$ and $T_\parallel^{(1)}$  read
\begin{align}\label{T1Cas}
&T_\perp^{(1)}=-\frac{8\pi A_{C}(\bm\Delta_0)}{\varepsilon_1\varepsilon_2 \varepsilon_3\varepsilon_4c_2\theta_{21}^2}\delta_{\mu_1\mu_2}\delta_{\mu_3\bar\mu_4}
\Big[\varepsilon_1(\varepsilon_3 \delta_{\mu_1\mu_3}-\varepsilon_4 \delta_{\mu_1\mu_4})(\bm s_{\mu_1}^*\cdot\bm\theta_{21})(\bm s_{\mu_1}\cdot \bm X_2)\nonumber\\
&+\varepsilon_2(\varepsilon_3 \delta_{\mu_1\bar\mu_3}-\varepsilon_4 \delta_{\mu_1\bar\mu_4})(\bm s_{\mu_1}\cdot\bm\theta_{21})(\bm s_{\mu_1}^*\cdot \bm X_2) \Big]\,,\nonumber\\
&T_\parallel^{(1)}=\frac{16\pi A_{C}(\bm\Delta_0)}{c_2^2}\delta_{\mu_1\mu_2}\delta_{\mu_3\bar\mu_4}\,(\bm\theta_{34}\cdot\bm\Delta_{0\perp})\,,\nonumber\\
&c_2=\varepsilon_1\varepsilon_2\theta_{21}^2+\varepsilon_3\varepsilon_4\theta_{34}^2\,,\quad
\bm X_2=\bm\Delta_{0\perp}-\frac{2\varepsilon_3\varepsilon_4}{c_2}(\bm\theta_{34}\cdot\bm\Delta_{0\perp})\bm\theta_{34}\,.
\end{align}
To derive this formula, we  use the  relation
\begin{align}\label{relation1}
i\eta\int d\bm\Delta_\perp A_{as}(\bm\Delta_\perp)\frac{(\bm\Delta_\perp-\bm q_{\perp})}{(\bm\Delta_\perp-\bm q)^2}=-\pi\bm q_{\perp}\,A_C(\bm q)\,,
\end{align}
 which is valid for any three-dimensional vector~$\bm q=q_\parallel\bm\nu+\bm q_{\perp}$, where $\bm\nu\cdot\bm q_{\perp}=0$ and $\bm\nu\cdot\bm\Delta_{\perp}=0$.

In the region  $\omega\ll \varepsilon_1$ and $\varepsilon_2 \theta_{21}\ll \min(m,\Delta_0)$, which provides the applicability of the Weizs\"acker-Williams approximation \cite{BLP1982}, the leading contribution to the amplitude of electroproduction is
\begin{align}\label{Tww}
&T_\perp^{(1)}=\frac{8\pi A_C(\bm\Delta_1)}{\omega^2\Delta_0^2 m^2}|\Gamma(1-i\eta)|^2\left(\frac{\xi_{W2}}{\xi_{W1}}\right)^{i\eta}
(\bm \Delta_{1\perp}\cdot {\bm T}_W)\,\delta_{\mu_1\mu_2}\,, \nonumber\\
&{\bm T}_W=\delta_{\mu_3\bar\mu_4} \big[(\varepsilon_3(\bm s_{\mu_3}\cdot\bm I_{W1})\,\bm s_{\mu_3}^* -\varepsilon_4
(\bm s_{\mu_4}\cdot\bm I_{W1})\,\bm s_{\mu_4}^*\big]
-\delta_{\mu_3\mu_4}\frac{m\omega\mu_3}{\sqrt{2}}I_{W0}\,\bm s_{\mu_3}^*\,,\nonumber\\
&\bm q_{W1}=-\frac{\varepsilon_3}{\omega}\bm \Delta_{0\perp}-\bm\zeta\,,\quad \bm q_{W2}=
-\frac{\varepsilon_4}{\omega}\bm \Delta_{0\perp}+\bm\zeta\,,\quad\bm\zeta=\frac{\varepsilon_3\varepsilon_4}{\omega}\bm\theta_{34}\,,\nonumber\\
&I_{W0}=(\xi_{W1}-\xi_{W2})F(x_W)+(\xi_{W1}+\xi_{W2}-1)(1-x_W)\frac{F'(x_W)}{i\eta}\,,\nonumber\\
&\bm I_{W1}=(\xi_{W1}\bm q_{W1}+\xi_{W2}\bm q_{W2})F(x_W)+(\xi_{W1}\bm q_{W1}-\xi_{W2}\bm q_{W2})(1-x_W)\frac{F'(x_W)}{i\eta}\,,\nonumber\\
&\xi_{W1}=\frac{m^2}{m^2+q_{W1}^2}\,,\quad \xi_{W2}=\frac{m^2}{m^2+q_{W2}^2}\,,\quad x_W=1-\frac{\Delta_{0\perp}^2\xi_{W1}\xi_{2W}}{m^2}\,,
\end{align}
where $\bm \Delta_{1\perp}=\varepsilon_2 \bm\theta_{21}$, $\Delta_{1\parallel}=m\omega/\varepsilon_1$, and the functions $F(x)$ and $F'(x)$ are defined in \eqref{T1Cnot}.
Note that the amplitude $T_{real}$ of $e^+e^-$ photoproduction by a real photon with the polarization vector $\bm e$ is \cite{LMS2012}
\begin{align}\label{Treal}
&T_{real}=\frac{8\pi\eta}{\omega\Delta_0^2 m^2}|\Gamma(1-i\eta)|^2 \left(\frac{\xi_{W2}}{\xi_{W1}}\right)^{i\eta}(\bm e\cdot {\bm T}_W)
\,,
\end{align}
where the function ${\bm T}_W$ is the same as in Eq.~\eqref{Tww}.

In the leading logarithmic approximation, the Coulomb corrections to the cross section in  the Coulomb field, proportional to $\ln^2(\varepsilon_1/m)$, are  originated from  three regions:
\begin{eqnarray}\label{reg2}
&&1.\, \dfrac{m^2}{\omega}\sim\Delta_{0\perp}\,,\quad \max\{\dfrac{m\omega}{\varepsilon_1},\Delta_{0\perp}\}\ll \Delta_{1\perp}\ll m \,,\nonumber\\
&&2.\,\max\{\Delta_{1\perp},\dfrac{m^2}{\omega}\}\ll\Delta_{0\perp}\ll m\,,\quad \dfrac{m\omega}{\varepsilon_1}\sim \Delta_{1\perp}\,,\nonumber\\
&&3.\,\Delta_0\sim m\,,\quad \dfrac{m\omega}{\varepsilon_1}\ll \Delta_{1\perp}\ll m \,,
\end{eqnarray}
where in all regions $ m\ll\omega\ll \varepsilon_1$ and $\zeta\sim m$.  Using  Eq.~\eqref{T1C}, we find  the amplitude $T_\perp^{(1)}$  in the first region in \eqref{reg2}:
\begin{align}\label{T1Cas1}
&T_\perp^{(1)}=-\frac{8\pi\varepsilon_1 A_C(\bm\Delta_0)\delta_{\mu_1\mu_2}}{\omega^2(m^2+\zeta^2)\Delta_{1\perp}^2}
{\cal M}_{as}\,,
\end{align}
where ${\cal M}_{as}$ is given in Eq.~\eqref{T1Cas11}. In the second region in \eqref{reg2}, the amplitude $T_\perp^{(1)}$  has the form
\begin{align}\label{T1Cas2}
&T_\perp^{(1)}=-\frac{8\pi\varepsilon_1 A_C(\bm\Delta_1)\delta_{\mu_1\mu_2}}{\omega^2(m^2+\zeta^2)\Delta_{0\perp}^2}
{\cal M}_{as}\,.
\end{align}
To derive Eqs.\eqref{T1Cas1} and \eqref{T1Cas2}, we use the relation \eqref{relation1}.
The expression for $T_\perp^{(1)}$ in  the third region in \eqref{reg2} is given  by Eq.~\eqref{Tww} with the replacement  $A_C(\Delta_1)\rightarrow A_{as}(\Delta_1)$.
In all regions in \eqref{reg2}, the terms $T_\perp^{(0)}$, $T_\parallel^{(0)}$, and $T_\parallel^{(1)}$  are suppressed   compared with $T_\perp^{(1)}$.

Performing  calculations with the logarithmic accuracy, we find that the  contributions of the first and second regions in \eqref{reg2} are equal to each other and two times smaller  than the contribution of the third region, so that the total result reads
\begin{align}\label{csCCl2tot}
&\sigma_{C}=-\frac{56\eta^2\alpha^2}{9\pi m^2}f(\eta)\ln^2\dfrac{\varepsilon_1}{m}\,,\quad f(\eta)=Re[\psi(1+i\eta)-\psi(1)]\,,
\end{align}
where $\psi(x)=d \Gamma(x)/dx$. However the origin of these contributions are different. The contribution of the third region can be easily obtained within   Weizs\"acker-Williams approximation. It corresponds to the Coulomb corrections to the  cross section of electroproduction in the Coulomb field by a relativistic particle noninteracting with this field \cite{IKSS1998,LM2000}. Therefore, the Coulomb corrections coming  from the third region   are given by the momentum transfer $\Delta_0\sim m$.
Using the language of  exchanges by the Coulomb quanta with the nucleus, we can say that the contributions to the Coulomb corrections of the first and second regions in  \eqref{reg2} correspond to the
case when all particles interact with the Coulomb center.  In this case, to derive the Coulomb corrections to the total cross section from Eqs.~\eqref{T1Cas1} and \eqref{T1Cas2}, it is necessary to use the relation \cite{LM2000,LMSS2005}
\begin{align}\label{relation2}
\int\,d\bm \Delta_\perp\Delta_{\perp}^2\left[|A_C(\bm\Delta)|^2-|A_B(\bm\Delta)|^2\right]=-32\pi^3\eta^2 f(\eta)\,.
\end{align}
Therefore, the Coulomb corrections coming  from the first and second regions   are given by the small momentum transfer $\Delta_0,\,\Delta_1\ll m$.

\section{Effect of screening}\label{Scr}
The effect of screening  is important if $\Delta_0\lesssim r_{scr}^{-1}$ or $\Delta_1\lesssim r_{scr}^{-1}$. In this case the main contribution to the integrals in (\ref{T1}) is given by
 $\Delta_\perp\sim\min(\Delta_0,\,\Delta_1)$. The effect of screening in the amplitude  $J_\lambda^{(1)}$ of photoproduction by a virtual photon
 \eqref{Jlambda} can be taken into account similar to the case of photoproduction by a real photon. Namely, one should multiply $J_\lambda^{(1)}$ in the Coulomb field by the atomic form factor $F(Q^2)$, where $\bm Q=\bm p_3+\bm p_4-\bm k$. This recipe is valid, because at $Q\ll m$  screening affect only the leading  in $\eta$ term of $J_\lambda^{(1)}$, while the Coulomb corrections to $J_\lambda^{(1)}$ are originated from the region  $Q\sim m$ where $F(Q^2)=1$.
Screening is taken into account in  the  matrix element  $j_\lambda$ of virtual photon bremsstrahlung  \eqref{jlambda}  via the function $A(\bm\Delta)$ in the atomic field, Eq.~\eqref{jlambda}.
Thus, the terms $T_\perp^{(0)}$ and $T_\parallel^{(0)}$ in the amplitude of electroproduction in the atomic field are  given directly by Eq.~\eqref{T0}. Then, multiplying the  integrands in Eq.~(\ref{T1C}) for the Coulomb field by the atomic form factor $F(Q^2)$ and making the replacement  $A_{as}(\bm\Delta_\perp)\rightarrow A(\bm\Delta_\perp)$, we obtain the terms $T_\perp^{(1)}$ and $T_\parallel^{(1)}$ in the atomic field. This  result will be valid for  any  values of  $\Delta_0$ and $\Delta_1$.

Let us now discuss  the impact of screening on the total cross section of electroproduction.
To obtain the result in the  leading logarithmic approximation, one should consider the region (cf. \eqref{reg1})
\begin{align}\label{reg1scr}
 m\ll\omega\ll \varepsilon_1\,,\quad \max\{\dfrac{m^2}{\omega},\,r_{scr}^{-1}\}\ll\Delta_{0\perp}\ll m\,,\quad \dfrac{m\omega}{\varepsilon_1}\ll \varepsilon_1\theta_{21}\ll m \,,\quad
\dfrac{\varepsilon_3\varepsilon_4}{\omega}\theta_{34}\sim m\,,
\end{align}
where the leading term of the cross section coincides with the Born result. Under the conditions \eqref{reg1scr},  the main contribution to the amplitude $T_B$ is given by the terms $T_{B\perp}^{(1)}$ \eqref{T1Cas11}, while the terms $T_{B\perp}^{(0)}$, $T_{B\parallel}^{(0)}$, and $T_{B\parallel}^{(1)}$  are suppressed. Performing the integration with the logarithmic accuracy, we obtain
for $\varepsilon_1> m^2r_{scr}\sim mZ^{-1/3}/\alpha$
\begin{align}\label{csl31}
\sigma=\frac{28\eta^2\alpha^2}{27\pi m^2}\left[\ln^3(m r_{scr})+3\ln\frac{\varepsilon_1}{m}\,\ln(m r_{scr})\ln\frac{\varepsilon_1}{m^2 r_{scr}}\right]\,.
\end{align}
 This result coincides with that obtained in Ref.~\cite{MUT}. For   $m\ll\varepsilon_1< m^2r_{scr}$ the total cross section in the leading logarithmic approximation is independent of $r_{scr}$ and coincides with Eq.~\eqref{csl3}.
Note that the asymptotics \eqref{csl31} has  good  accuracy only at very high energy. To obtain the result with high accuracy, it is necessary to perform the integration of our results  beyond the leading logarithmic approximation.

In the leading logarithmic approximation, the  Coulomb corrections  to the  total cross section of electroproduction in  the atomic field are  originated from  three regions (cf. \eqref{reg2}):
\begin{align}\label{reg3}
&1.\, \max\{\dfrac{m^2}{\omega},r_{scr}^{-1}\}\sim\Delta_{0\perp}\,,\quad \max\{\dfrac{m\omega}{\varepsilon_1},\Delta_{0\perp}\}\ll \Delta_{1\perp}\ll m \,,\nonumber\\
&2.\,\max\{\Delta_{1\perp},\dfrac{m^2}{\omega}\}\ll\Delta_{0\perp}\ll m\,,\quad \max\{\dfrac{m\omega}{\varepsilon_1},\,r_{scr}^{-1}\}\sim \Delta_{1\perp}\,,\nonumber\\
&3.\,\Delta_0\sim m\,,\quad \dfrac{m\omega}{\varepsilon_1}\ll\Delta_{1\perp}\ll m \,,
\end{align}
where in all regions $ m\ll\omega\ll \varepsilon_1$ and  $\zeta=\varepsilon_3\varepsilon_4\theta_{34}/\omega\sim m$.

To obtain the amplitude $T_\perp^{(1)}$  in the first region in \eqref{reg3}, we use the following transformation
\begin{align}\label{relation3}
&i\eta\int d\bm\Delta_\perp A(\bm\Delta_\perp)F\left((\bm q-\bm\Delta_\perp)^2\right)\frac{(\bm q_{\perp}-\bm\Delta_\perp)}{(\bm q-\bm\Delta_\perp)^2}\nonumber\\
&=-\frac{1}{4\pi}\int d\bm\Delta_\perp d\bm r   A(\bm\Delta_\perp)\bm\nabla_\perp V(\bm r)e^{-i (\bm q-\bm \Delta_\perp)\bm r}\nonumber\\
&=-i\pi\int d\bm r \, \bm{\nabla_\perp V(r)}\exp[-i\chi(\rho)-i\bm q\cdot\bm r]=\pi\bm q_{\perp}\,A(\bm q)\,,
\end{align}
where the function $A(\bm\Delta)$ is given in Eq.~\eqref{jlambda}. Then the amplitude $T_\perp^{(1)}$  in the first region in \eqref{reg3} is
\begin{align}\label{T1Cas3}
&T_\perp^{(1)}=-\frac{8\pi\varepsilon_1 A(\bm\Delta_0)\delta_{\mu_1\mu_2}}{\omega^2(m^2+\zeta^2)\Delta_{1\perp}^2}
{\cal M}_{as}\,,
\end{align}
where ${\cal M}_{as}$ is given in Eq.~\eqref{T1Cas1}. Similar  to Eq.~\eqref{T1Cas3},  we obtain  the amplitude $T_\perp^{(1)}$  in the second region in \eqref{reg3}:
\begin{align}\label{T1Cas4}
&T_\perp^{(1)}=-\frac{8\pi\varepsilon_1 A_1(\bm\Delta_1)\delta_{\mu_1\mu_2}}{\omega^2(m^2+\zeta^2)\Delta_{0\perp}^2}
{\cal M}_{as}\,,\\\nonumber
&A_1(\bm\Delta)=-\frac{i}{\Delta_{\perp}^2}\int d\bm r\,\exp[-i\bm\Delta\cdot\bm r-i\chi(\rho)]\bm\Delta_{\perp}\cdot\bm\nabla_\perp V_C(r)\,.
\end{align}
The amplitude  $T_\perp^{(1)}$ in  the third region in \eqref{reg3} is independent of $r_{scr}$ and coincides with the expression \eqref{Tww} with the replacement
 $A_C(\Delta_1)\rightarrow A_{as}(\Delta_1)$, Eq.~\eqref{ACJCas}. In all regions in \eqref{reg3}, the terms $T_\perp^{(0)}$, $T_\parallel^{(0)}$, and $T_\parallel^{(1)}$  are suppressed   compared with $T_\perp^{(1)}$.

For the total cross section, the contributions of the first and second regions to  the Coulomb corrections calculated within logarithmic accuracy are equal to each other,
\begin{align}\label{csCCl2tot1}
&\delta\sigma_{C}^{(1)}=\delta\sigma_{C}^{(2)}= -\frac{56\eta^2\alpha^2}{9\pi m^2}f(\eta)\ln(mr_{csr})\ln\dfrac{\varepsilon_1}{m^2 r_{csr}}
\end{align}
for  $\varepsilon_1>m^3 r_{scr}^2$  and
\begin{align}\label{csCCl2tot2}
&\delta\sigma_{C}^{(1)}=\delta\sigma_{C}^{(2)}=-\frac{14\eta^2\alpha^2}{9\pi m^2}f(\eta)\ln^2\dfrac{\varepsilon_1}{m}\,
\end{align}
for  $m\ll\varepsilon_1<m^3 r_{scr}^2$.
To derive Eqs.~\eqref{csCCl2tot1} and  \eqref{csCCl2tot2} we use  the relations \cite{LMSS2005}
\begin{align}\label{relation2}
&\int\,d\bm \Delta_{0\perp}\Delta_{0\perp}^2\left[|A(\bm\Delta_0)|^2-|A_B(\bm\Delta_0)|^2\right]=-32\pi^3\eta^2 f(\eta)\,,\nonumber\\
&\int\,d\bm \Delta_{1\perp}\Delta_{1\perp}^2\left[|A_1(\bm\Delta_1)|^2-|A_{1B}(\bm\Delta_1)|^2\right]=-32\pi^3\eta^2 f(\eta)\,,\nonumber\\
&A_{1B}(\bm\Delta_1)= -4\pi\eta /\Delta_1^2\,,
\end{align}
valid for any atomic potential $V(r)$. The contribution of the  third region is independent of $r_{scr}$ and equals
\begin{align}\label{csCCl2tot3}
&\delta\sigma_{C}^{(3)}=-\frac{28\eta^2\alpha^2}{9\pi m^2}f(\eta)\ln^2\dfrac{\varepsilon_1}{m}\,
\end{align}
for all $\varepsilon_1\gg m$. Thus, the total Coulomb corrections $\sigma_{C}=\delta\sigma_{C}^{(1)}+\delta\sigma_{C}^{(2)}+\delta\sigma_{C}^{(3)}$  are
\begin{align}\label{csCCl2tot4}
&\sigma_{C}=-\frac{28\eta^2\alpha^2}{9\pi m^2}f(\eta)\left[\ln^2\dfrac{\varepsilon_1}{m}+4\ln(mr_{csr})\ln\dfrac{\varepsilon_1}{m^2 r_{csr}}\right]
\end{align}
for $\varepsilon_1>m^3 r_{scr}^2$ and
\begin{align}\label{csCCl2tot5}
&\sigma_{C}=-\frac{56\eta^2\alpha^2}{9\pi m^2}f(\eta)\ln^2\dfrac{\varepsilon_1}{m}\,
\end{align}
for $m\ll\varepsilon_1<m^3 r_{scr}^2$, see Eq.~\eqref{csCCl2tot}. Note that the contribution $\delta\sigma_{C}^{(1)}+\delta\sigma_{C}^{(2)}$, coming from small momentum transfers $\Delta_0\ll m$, approximately equals to $\delta\sigma_{C}^{(3)}$, coming from $\Delta_0\sim m$, up to very high energy $\varepsilon_1$.

\section{Impact of the Coulomb corrections on the differential cross section}
To compare the exact in $\eta$ differential cross section with the Born result, we introduce the dimensionless quantity S,
\begin{equation}\label{S}
S=\sum_{\mu_1\mu_2\mu_3\mu_4}\Bigg|\frac{\varepsilon_1 m^4 {\cal T}_{\mu_1\mu_2\mu_3\mu_4}}{\eta (2\pi)^2}\Bigg|^2 \,,
\end{equation}
 which is the normalized differential cross section summed over polarizations of all particles. We  consider the region $\Delta_{0\parallel}\gg 1/r_{scr}$ and $\omega/\varepsilon_1\gg 1/(mr_{scr})$, where it is not necessary to take screening into account. Besides, for  $\Delta_{0\perp}\gg \Delta_{0\parallel}$  the function $S$ depends only on the ratios $\varepsilon_i/\varepsilon_1$, but not on the energy $\varepsilon_1$ itself.

\begin{figure}
\centering
\includegraphics[width=0.7\linewidth]{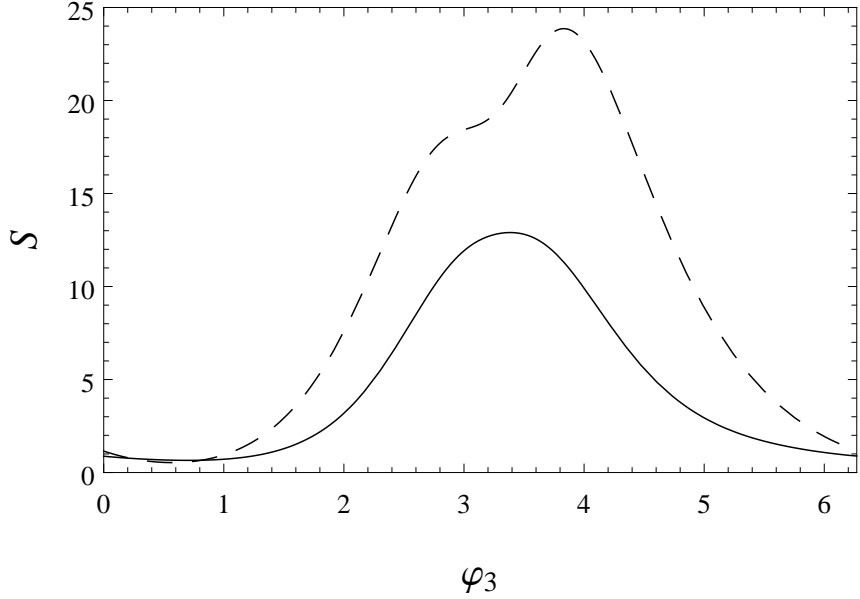}
\caption{The quantity $S$, see Eq.~\eqref{S}, as a function of the azimuth  angle $\varphi_3$ for  $\varepsilon_2/\varepsilon_1=0.28$, $\varepsilon_3/\varepsilon_1=0.42$, $\varepsilon_4/\varepsilon_1=0.3$, $p_{2\perp}=0.3 m$, $p_{3\perp}=0.5 m$,  $p_{4\perp}=1.2m$,  and $\varphi_2=\pi/4$; the Born result (dashed curve) and the exact in $\eta$ result for  $\eta=0.6$
(solid curve).}
\label{fig:cs1}
\end{figure}

  We direct  $z$-axis along $\bm p_1$ and $x$-axis along $\bm p_{4\perp}$.
In Fig.~\ref{fig:cs1} we show the dependence of $S$ on  $\varphi_3$  at some values of $\varepsilon_i$, $p_{i\perp}$, and $\varphi_2$, where  $\varphi_2$ is the azimuth  angle of  $\bm p_{2\perp}$
and  $\varphi_3$ is the azimuth angle of  $\bm p_{3\perp}$.

\begin{figure}
\centering
\includegraphics[width=1.\linewidth]{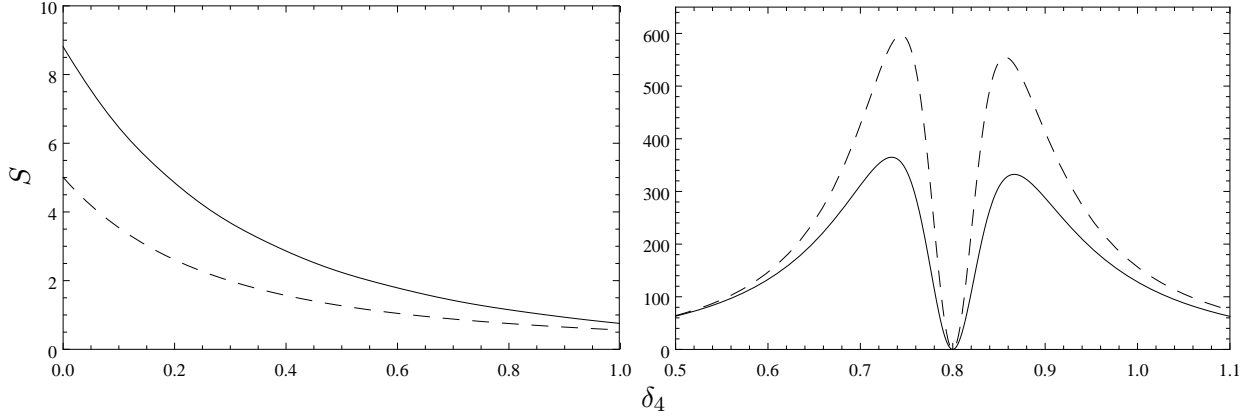}
\caption{The quantity $S$, see Eq. \eqref{S}, as a function of $\delta_4=p_{4\perp}/m$ for $\varepsilon_1=100m$, $\varepsilon_2/\varepsilon_1=0.28$, $\varepsilon_3/\varepsilon_1=0.42$, $\varepsilon_4/\varepsilon_1=0.3$, $p_{2\perp}=0.3 m$, $p_{3\perp}=0.5 m$,  $\varphi_2=\varphi_3=0$ (left  picture), and $\varphi_2=\varphi_3=\pi$ (right  picture); Born result (dashed curve) and the exact in $\eta$ result for $\eta=0.6$ (solid curve).}
\label{fig:cs2}
\end{figure}

In Fig.~\ref{fig:cs2} the dependence of $S$ on $\delta_4=p_{4\perp}/m$ is shown at some values of $\varepsilon_i$ and $\bm p_{i\perp}$. In the right picture in Fig.~\ref{fig:cs2} the point $\delta_4=0.8$,
where $S=0$, corresponds to the momentum transfer $\Delta_{0\perp}=0$. It is seen from Figs.~\ref{fig:cs1} and \ref{fig:cs2} that the Coulomb corrections significantly modify  the differential cross section  compared with the Born result.

\begin{figure}
	\centering
	\includegraphics[width=0.7\linewidth]{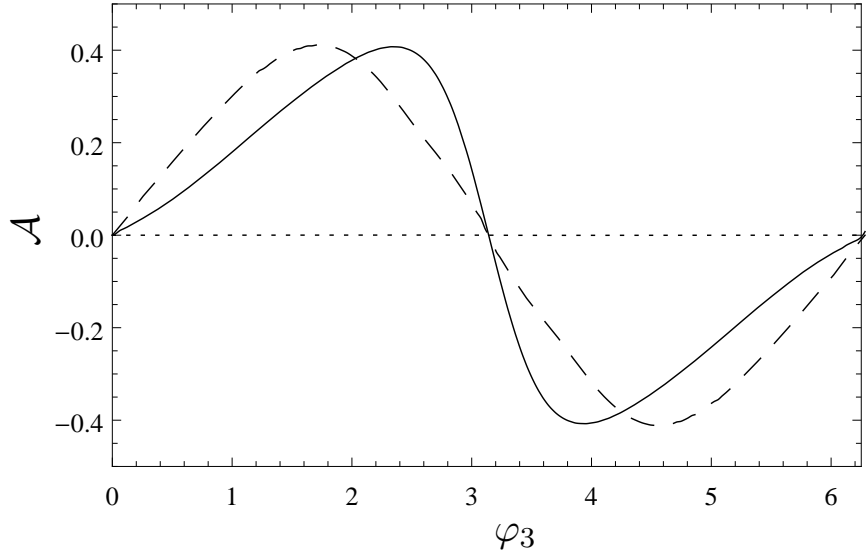}
	\caption{The quantity $\mathcal{A}$, see Eq. \eqref{as1},  as a function of $\varphi_3$ for  $\varepsilon_2/\varepsilon_1=0.28$,  $\varepsilon_3/\varepsilon_1=0.42$, $\varepsilon_4/\varepsilon_1=0.3$, $p_{2\perp}=0.3 m$, $p_{3\perp}=0.5 m$, $p_{4\perp}=1.2m$, $\eta=0.6$;  $\varphi_2=0$ (solid curve) and  $\varphi_2=\pi$ (dashed curve). In the Born approximation $\mathcal{A}=0$ (dotted curve).}
	\label{fig:as}
\end{figure}

The exact in $\eta$ differential cross section for the polarized incoming particle possesses the azimuth  asymmetry  $\mathcal{A}$,
\begin{align}\label{as1}
\mathcal{A}=\frac{S_+-S_-}{S_++S_-}\,,\quad
S_{\pm}=\sum_{\mu_2\mu_3\mu_4}\Bigg|\frac{\varepsilon_1 m^4 {\cal T}_{\pm\mu_2\mu_3\mu_4}}{\eta (2\pi)^2}\Bigg|^2 \,.
\end{align}
In the Born approximation the asymmetry vanishes for any $\bm p_i$ due to the  relation
\begin{equation}
{\cal T}^B_{\mu_1\mu_{2}\mu_{3}\mu_{4}}=-\mu_{1}\mu_{2}\mu_{3}\mu_{4}\left({\cal T}^B_{\overline{\mu}_{1}\overline{\mu}_{2}\overline{\mu}_{3}\overline{\mu}_{4}}\right)^*\,,
\end{equation}
following  from  Eq.~\eqref{T1CB}. However,  this relation is not valid for the Coulomb corrections because the integrand in Eq. \eqref{T1C} is not a real quantity.
The asymmetry     $\mathcal{A}$  is shown in Fig.~\ref{fig:as}  as a function of  $\varphi_3$  at some values of $\varepsilon_i$, $p_{i\perp}$, and $\varphi_2$. As it should be, the asymmetry vanishes when all momenta are in the same plane ($\varphi_2=0,\pi$ and $\varphi_3=0,\pi$ in Fig. \ref{fig:as}). It is seen that the asymmetry can reach tens of percent.

\section{Conclusion}
Using the quasiclassical approximation, we have derived the exact in the parameter $\eta$ differential cross section of high-energy electroproduction in the atomic field. The helicity amplitudes of the process for the Coulomb field are given in Eqs.~\eqref{T0} and \eqref{T1C}, and the modification of these formulas for the atomic field (the effect of screening) are pointed out in Sec. \ref{Scr}.
The Coulomb corrections substantially modify  the differential cross section compared with the Born result. For the polarized incoming electron, the Coulomb corrections lead to the azimuth asymmetry in the differential cross section, Eq.~\eqref{as1}.   The leading logarithmic contribution to the Coulomb corrections to the total cross section is given  not only by  moderate momentum transfers $\Delta_0\sim m$, but also by  small momentum transfers  $\Delta_0\ll m$. Emphasize that the latter contribution appears due to interaction of the incoming electron with the atomic field.

\section*{Acknowledgement}
We are grateful to R.N. Lee for important discussions. This work has been supported by Russian Science Foundation (Project No. 14-50-00080). It has been also supported in part by
RFBR (Grant No. 16-02-00103).


\begin{thebibliography}{99}

\bibitem{A1} A1 Collaboration (H. Merkel et al.) Phys.Rev.Lett. {\bf106}, 251802 (2011).

\bibitem{APEX} APEX Collaboration (S. Abrahamyan et al.) Phys.Rev.Lett. {\bf 107}, 191804 (2011).

\bibitem{Bhabha1} H. Bhabha, Proc. Cambridge Phil. Soc. {\bf31}, 394 (1935).

\bibitem{Bhabha2} H. Bhabha, Proc. Roy. Soc. (London) {\bf A152}, 559 (1935).

\bibitem{Racah36} G. Racah, Nuovo Cimento {\bf4}, 66 (1936).

\bibitem{Racah37} G. Racah, Nuovo Cimento {\bf4}, 112 (1937).

\bibitem{BLP1982} V. B. Berestetski, E. M. Lifshits, and L. P. Pitayevsky, {\it Quantum Electrodynamics} (Pergamon, Oxford, 1982).

\bibitem{BKW} M. M. Block, D. T. King, and M. Wada, Phys. Rev. {\bf96}, 1627 (1954).

\bibitem{MU} T. Murota and A. Veda, Progr. Theoret. Phys. (Kyoto) {\bf16}, 497 (1956).

\bibitem{MUT} T. Murota, A. Veda, and H. Tanaka, Progr. Theoret. Phys.(Kyoto) {\bf16}, 482 (1956).

\bibitem{Johnson}E. G. Johnson Phys.Rev {\bf 140}, 1005 (1965).

\bibitem{Brodsky} S. Brodsky and S. Ting, Phys. Rev. {\bf 145}, 1018 (1966).

\bibitem{BjCh} J. D. Bjorken and M. Chen, Phys. Rev. {\bf 154}, 1335 (1967).

\bibitem{Henry} R. Henry, Phys. Rev. {\bf154}, 1534 (1967).

\bibitem{Homma} S. Homma, A . Itano, K . Nishikawa, and M. Hayashi, Proc. Phys. Soc. Japan {\bf3}, 1230 (1974).

\bibitem{BM1954} H.~A. Bethe and L.~C. Maximon, Phys. Rev. {\bf 93}, 768 (1954).

\bibitem{DBM1954} H.~Davies, H.~A. Bethe, and L.~C. Maximon, Phys. Rev. {\bf 93}, 788 (1954).

\bibitem{OlsenMW1957} H.~A. Olsen, L.~C.  Maximon, and H. Wergeland, Phys. Rev.  {\bf 106}, 27 (1957).

\bibitem{Fu} W. Furry, Phys. Rev. {\bf 46}, 391 (1934).

\bibitem{ZM}  A. Sommerfeld, A. Maue, Ann. Phys. {\bf 22}, 629 (1935).

\bibitem{LMS00} R.~N. Lee, A.~I. Milstein, V. M. Strakhovenko, Zh. Eksp. Teor. Fiz. {\bf 117}, 75 (2000) [JETP {\bf 90}, 66 (2000)].

\bibitem{KM2015} P.~A. Krachkov and  A.~I. Milstein, Phys. Rev. A  {\bf 91}, 032106  (2015).

\bibitem{LMSS2005} R.~N. Lee, A.~I. Milstein,  V.~M. Strakhovenko, and O.~Ya. Schwarz, Zh. Eksp. Teor. Fiz. {\bf 127}, 5 (2005) [JETP {\bf 100}, 1 (2005)].

\bibitem{LMS2012} R.~N. Lee, A.~I. Milstein, and V.~M. Strakhovenko, Phys. Rev. A {\bf 85}, 042104 (2012).

\bibitem{KLM2014} P.~A. Krachkov, R.~N. Lee, and  A.~I. Milstein, Phys. Rev. A {\bf 90}, 062112 (2014).

\bibitem{KLM2015d} P. A. Krachkov, R.N. Lee,  and A. I. Milstein, Phys. Rev. A {\bf 91}, 062109 (2015).

\bibitem{IKSS1998} D. Yu. Ivanov, E. A. Kuraev, A. Schiller, and V. G. Serbo, Phys. Lett. B {\bf 442}, 453 (1998).

\bibitem{LM2000} R.N. Lee, A.I. Milstein, Phys. Rev. A {\bf 61},  032103  (2000).

\end{thebibliography}
 \end{document}